\documentclass[apjl]{emulateapj}
\usepackage{psfig,amsfonts,amsmath,graphicx,natbib,apjfonts}
\def\nod{\nodata}

\def\to{$\rightarrow$}

\def\cfa{1}
\def\ucb{2}
\def\steward{3}
\def\noao{4}
\def\udel{5}
\def\csic{6}
\def\ucf{7}
\def\hcmiu{8}
\def\mcgill{9}

\begin{document}

\title{Simultaneous Multi-Wavelength Observations of Magnetic Activity
in Ultracool Dwarfs. III.  X-ray, Radio, and H$\alpha$ Activity Trends
in M and L Dwarfs}

\author{
E.~Berger\altaffilmark{\cfa},
G.~Basri\altaffilmark{\ucb},
T.~A.~Fleming\altaffilmark{\steward},
M.~S.~Giampapa\altaffilmark{\noao},
J.~E.~Gizis\altaffilmark{\udel},
J.~Liebert\altaffilmark{\steward},
E.~L.~Mart{\'{\i}}n\altaffilmark{\csic,}\altaffilmark{\ucf},
N.~Phan-Bao\altaffilmark{\hcmiu},
and R.~E.~Rutledge\altaffilmark{\mcgill}
}

\altaffiltext{\cfa}{Harvard-Smithsonian Center for Astrophysics, 60
Garden Street, Cambridge, MA 02138}

\altaffiltext{\ucb}{Astronomy Department, University of
California, Berkeley, CA 94720}

\altaffiltext{\steward}{Department of Astronomy and Steward
Observatory, University of Arizona, 933 North Cherry Avenue,
Tucson, AZ 85721}

\altaffiltext{\noao}{National Solar Observatory, National Optical
Astronomy Observatories, Tucson, AZ 85726}

\altaffiltext{\udel}{Department of Physics and Astronomy,
University of Delaware, Newark, DE 19716}

\altaffiltext{\csic}{CSIC-INTA Centro de Astrobiologia, Torrejon 
de Ardoz, Madrid, Spain}

\altaffiltext{\ucf}{University of Central Florida, Department of
Physics, PO Box 162385, Orlando, FL 32816}

\altaffiltext{\hcmiu}{Department of Physics, HCMIU, Vietnam National
University Administrative Building, Block 6, Linh Trung Ward, Thu Duc
District, HCM, Vietnam}

\altaffiltext{\mcgill}{Department of Physics, McGill University,
3600 University Street, Montreal, QC H3A 2T8, Canada}

\begin{abstract} 
As part of our on-going investigation into the magnetic field
properties of ultracool dwarfs, we present simultaneous radio, X-ray,
and H$\alpha$ observations of three M9.5-L2.5 dwarfs (BRI\,0021-0214,
LSR\,060230.4+391059, and 2MASS J052338.2$-$140302).  We do not detect
X-ray or radio emission from any of the three sources, despite
previous detections of radio emission from BRI\,0021 and 2M0523$-$14.
Steady and variable H$\alpha$ emission are detected from 2M0523$-$14
and BRI\,0021, respectively, while no H$\alpha$ emission is detected
from LSR\,0602+39.  Overall, our survey of nine M8-L5 dwarfs doubles
the number of ultracool dwarfs observed in X-rays, and triples the
number of L dwarfs, providing in addition the deepest limits to date,
${\rm log}(L_X/L_{\rm bol})\lesssim -5$.  With this larger sample we
find the first clear evidence for a substantial reduction in X-ray
activity, by about two orders of magnitude, from mid-M to mid-L
dwarfs.  We find that the decline in both X-rays and H$\alpha$ roughly
follows $L_{X,{\rm H\alpha}}/L_{\rm bol}\propto 10^{-0.4\times({\rm
SP-M6})}$ for ${\rm SP\gtrsim M6}$.  In the radio band, however, the
luminosity remains relatively unchanged from M0 to L4, leading to a
substantial increase in $L_{\rm rad}/L_{\rm bol}$.  Our survey also
provides the first comprehensive set of simultaneous
radio/X-ray/H$\alpha$ observations of ultracool dwarfs, and reveals a
clear breakdown of the radio/X-ray correlation beyond spectral type
M7, evolving smoothly from $L_{\rm \nu,rad}/L_X\approx 10^{-15.5}$ to
$\sim 10^{-11.5}$ Hz$^{-1}$ over the narrow spectral type range M7-M9.
This breakdown reflects the substantial reduction in X-ray activity
beyond M7, but its physical origin remains unclear since, as evidenced
by the uniform radio emission, there is no drop in the field
dissipation and particle acceleration efficiency.  Based on the
results of our survey, we conclude that a further investigation of
magnetic activity in ultracool dwarfs will benefit from a two-pronged
approach: multi-rotation observations of nearby known active sources,
and a snapshot survey of a large sample within $\sim 50$ pc to uncover
rare flaring objects.
\end{abstract}
 

\section{Introduction}
\label{sec:intro}

The study of magnetic fields in fully convective low mass stars and
brown dwarfs (hereafter, ultracool dwarfs) has progressed rapidly in
recent years thanks to a broad range of observational techniques.
These include observations of magnetic activity indicators such as
radio, X-ray, and H$\alpha$ emission (e.g.,
\citealt{rbm+00,whw+04,ber06}), Zeeman broadening of FeH molecular
lines in Stokes $I$ \citep{rb07}, and time-resolved spectropolarimetry
in Stokes $V$ (Zeeman-Doppler imaging; \citealt{dmp+08,mdp+08}).  Each
of these techniques is sensitive to specific properties of the
magnetic field.  Zeeman broadening provides a measure of the {\it
integrated} surface magnetic flux, $Bf$ \citep{rb06}, where $B$ is the
magnetic field strength and $f$ is the field covering fraction.
Zeeman-Doppler imaging (ZDI), on the other hand, only allows for a
reconstruction of the large-scale (multipole number of $\ell\lesssim
{\rm few}$) surface field, but it also provides information on the
field {\it topology} \citep{dmp+08,mdp+08}.

The ZDI technique has led to a characterization of surface fields in
several M dwarfs, revealing a transition from mainly toroidal and
non-axisymmetric fields in M0-M3 objects to predominantly poloidal
axisymmetric fields in mid-M dwarfs \citep{dmp+08,mdp+08}.  The
apparent shift in field geometry coincides with the transition to full
convection, and may reflect a change in dynamo mechanism.  The field
strengths inferred from ZDI are $\sim 0.1-0.8$ kG, while those from
the FeH Zeeman broadening method range from a non-detectable level of
$\lesssim 0.1$ kG up to $\sim 3$ kG \citep{rb07}.  However, in the few
overlapping cases, the two techniques result in substantially
different field strengths\footnotemark\footnotetext{The FeH Zeeman
broadening technique leads to fields of 3.9 kG for EV Lac, 2.9 kG for
AD Leo, and $>3.9$ kG for YZ CMi \citep{rb07}, while the ZDI technique
leads to much weaker fields of $0.5-0.6$ kG, 0.2 kG, and 0.55 kG for
the three objects, respectively \citep{mdp+08}.}, most likely as a
result of sensitivity to different physical scales \citep{rb09}.

Despite their potential for measuring surface fields, neither of the
two Zeeman techniques provide information on whether or how the
magnetic fields couple to the stellar atmosphere and dissipate their
energy.  In addition, both techniques are challenging for the much
dimmer L dwarfs since high signal-to-noise ratios of $\gtrsim 100$
are required, and they also lose sensitivity at high rotation velocity
due to line broadening \citep{rb07}.  Finally, since only the large
scale or integrated field is traced with these observations, they are
unlikely to reveal field variability on short timescale and small
physical scales due to flares.

These issues can be overcome through observations of various radiative
processes that trace plasma heating and particle acceleration by the
magnetic field.  Of particular interest are chromospheric and coronal
emission, traced by H$\alpha$ and X-rays, respectively, as well as
radio emission from accelerated electrons (gyrosynchrotron or coherent
emission).  These diagnostics not only provide an independent test for
the existence, strength, and geometry of the magnetic fields, but they
also allow us to infer their dissipation, particularly on short
timescales and small physical scales.  These techniques are not biased
against rapid rotation and can be applied much more easily in late-M,
L, and T dwarfs.

In this context, large samples of ultracool dwarfs have been observed
in H$\alpha$, primarily as a by-product of spectroscopic observations
aimed at measurements of stellar properties (e.g.,
\citealt{gmr+00,mb03,whw+04,rb07}).  These observations reveal a
saturated level of emission in the early M dwarfs, with $L_{\rm
H\alpha}/L_{\rm bol}\approx 10^{-3.5}$, followed by a rapid decline
beyond spectral type M7 \citep{gmr+00,whw+04}.  The H$\alpha$
rotation-activity relation that exists in higher mass stars, breaks
down at about the same spectral type \citep{mb03}.  In the radio band,
more than 100 ultracool dwarfs have been observed, revealing the
opposite trend; namely, $L_{\rm rad}/L_{\rm bol}$ appears to increase
with later spectral type, at least to $\sim {\rm L3}$ \citep{ber06},
and a rotation-activity relation may exist even beyond M7
\citep{bbg+08}.  In X-rays only a few objects beyond M7 have been
observed (e.g., \citealt{fgs+93,nbc+99,fgs+00,rbm+00,bbg+08,rs08}),
revealing a possible minor decline from the saturated level of
$L_X/L_{\rm bol}\approx 10^{-3}$ that is observed in M0-M6 dwarfs.  A
breakdown of the rotation-activity relation in X-rays has also been
proposed \citep{bbg+08}.

Still, few attempts have been made to study the correlated behavior of
magnetic activity indicators in ultracool dwarfs.  Such studies are of
prime importance because they can directly link magnetic dissipation
and particle acceleration with chromospheric and coronal heating,
particularly in the context of time-variable phenomena (flares).
These studies can also test whether the radio/X-ray and
X-ray/H$\alpha$ correlations that exist in higher mass stars are also
observed in ultracool dwarfs.  Such studies are observationally
challenging, since they require simultaneous coverage with a wide
range of facilities.  Despite this difficulty, ultracool dwarfs are
particularly well-suited to these observations since their rapid
rotation allows us to cover multiple rotations in several hours.

Over the past few years we have carried out such observations (radio,
H$\alpha$, X-rays) for several ultracool dwarfs.  Results for three
late-M dwarfs were published in \citet{bgg+08} (hereafter, Paper I)
and \citet{bbg+08} (hereafter, Paper II), while observations of three
L dwarfs were published in \citet{brr+05} and \citet{brp+09}.  In the
course of this survey we have found an overall lack of temporal
correlation between the various activity indicators.  In particular,
the tight radio/X-ray correlation breaks down at about spectral type
M7, with later objects exhibiting suppressed X-ray emission.  This
result may be tied to coronal stripping by rapid rotation, which is
ubiquitous beyond M7 (Paper II) or to reduced coronal densities, which
would strongly suppress X-ray bremsstrahlung emission.  Moreover, we
have uncovered several cases of rotationally-modulated emission in
both H$\alpha$ and radio that arise from large-scale magnetic hot
spots (Paper I and \citealt{brp+09}).

In this paper we present new simultaneous multi-wavelength
observations of three ultracool dwarfs (M9.5, L1, and L2.5).  Taken
together with the previous observations, our program doubles the
number of ultracool dwarfs, and triples the number of L dwarfs,
observed in X-rays.  Using this extended sample, and taking advantage
of the simultaneous nature of the observations, we investigate in this
paper the overall activity trends of ultracool dwarfs.

\section{Targets and Observations}
\label{sec:obs}

We present observations of three nearby ultracool dwarfs,
BRI\,0021-0214 (M9.5; hereafter, BRI0021), LSR\,060230.4+391059 (L1;
hereafter, LSR0602+39), and 2MASS J052338.2$-$140302 (L2.5; hereafter,
2M0523$-$14).  The properties of these sources, including the results
of previous H$\alpha$ and radio observations are summarized in
Table~\ref{tab:sources}.  Both BRI0021 and 2M0523$-$14 have been shown
to exhibit variable radio emission \citep{ber02,adh+07}, while
LSR0602+39 was not previously detected \citep{ber06}.  Similarly,
BRI0021 exhibited variable H$\alpha$ emission in the past, while
2M0523$-$14 appears to have steady H$\alpha$ emission, and LSR0602+39
shows no detectable emission.  Finally, of the three sources only
BRI0021 has been observed in X-rays (with ROSAT) and was not detected
to a limit of $L_X/L_{\rm bol}\lesssim -4.7$ \citep{nbc+99}.

The observations presented here took place on 2007 December 30
(2M0523$-$14), 2008 January 18 (LSR0602+39), and 2008 October 29
(BRI0021).

\subsection{Radio} 
\label{sec:rad}

Very Large Array\footnotemark\footnotetext{The VLA is operated by the
National Radio Astronomy Observatory, a facility of the National
Science Foundation operated under cooperative agreement by Associated
Universities, Inc.} observations were obtained simultaneously at 4.86
and 8.46 GHz in the standard continuum mode ($2\times 50$ MHz
contiguous bands), with thirteen antennas used at each frequency.  The
observational setup for each source is summarized in
Table~\ref{tab:radio}.  Data reduction and analysis followed the
procedures outlined in Papers I and II.  We did not detect radio
emission from any of the three sources.  The upper limits on the flux
and $L_{\rm rad}/L_{\rm bol}$ are listed in Table~\ref{tab:radio}.

\subsection{X-Rays}
\label{sec:xrays}

Observations were performed with the Chandra/ACIS-I2
frontside-illuminated chip.  A total of 29.80, 29.66, and 29.77 ks
were obtained for BRI0021, LSR0602+39, and 2M0523$-$14, respectively.
The data were analyzed using CIAO version 3.4, and counts were
extracted in a $2''$ radius circle centered on the position of each
source.

No X-ray emission was detected from any of the three sources above the
background level of about 1.5 counts per 30 ks.  Thus, the resulting
upper limits are about 7 counts ($95\%$ confidence level).  To convert
to a limit on the X-ray flux and $L_X/L_{\rm bol}$ we use an energy
conversion factor of $1\,{\rm cps}=3.4\times 10^{-12}$ erg cm$^{-2}$
s$^{-1}$, appropriate for a 1 keV Raymond-Smith plasma model in the
$0.2-2$ keV range.  The resulting limits are provided in
Table~\ref{tab:xray}.

\subsection{Optical Spectroscopy and H$\alpha$ Imaging} 
\label{sec:optical}

We observed\footnotemark\footnotetext{Program GS-2008A-Q-79.}
2M0523$-$14 with the Gemini Multi-Object Spectrograph (GMOS;
\citealt{hja+04}) mounted on the Gemini-South 8-m telescope.  The
observations were performed with the B600 grating set at a central
wavelength of 5250 \AA, and with a $1''$ slit.  The individual 300 s
exposures were reduced using the {\tt gemini} package in IRAF (bias
subtraction, flat-fielding, and sky subtraction).  Wavelength
calibration was performed using CuAr arc lamps.  The spectra cover
$3840-6680$ \AA\ at a resolution of about 5 \AA.  A series of 73
exposures were obtained with an on-source efficiency of $94\%$.  A
summary of the observations and results is provided in
Table~\ref{tab:ha}.  We find non-variable H$\alpha$ emission with a
mean equivalent width of $1.00\pm 0.05$ \AA, about a factor of three
time higher than previously observed by \citet{rb08}.  This
corresponds to a ratio of ${\rm log}(L_{\rm H\alpha}/L_{\rm
bol})=-6.05$.

We observed LSR0602+39 with the Dual Imaging spectrograph (DIS)
mounted on the ARC 3.5-m telescope at Apache Point Observatory.  The
observations were performed with the R1200 grating set at a central
wavelength of 6980 \AA, and with a $1.5''$ slit.  The individual 600 s
exposures were reduced using standard packages in IRAF, and wavelength
calibration was performed using HeNeAr arc lamps.  The spectra cover
$6400-7550$ \AA\ at a resolution of about 1.3 \AA.  A series of 38
exposures were obtained with an on-source efficiency of $93\%$.  A
summary of the observations and results is provided in
Table~\ref{tab:ha}.  No H$\alpha$ emission is detected in any of the
individual spectra, or in the combined spectrum to a limit of $<0.25$
\AA.  This corresponds to a ratio of ${\rm log}(L_{\rm H\alpha}/L_{\rm
bol})<-6.3$, a factor of 2 times lower than the limit found by
\citet{rb08}.

\begin{figure}
\centerline{\psfig{file=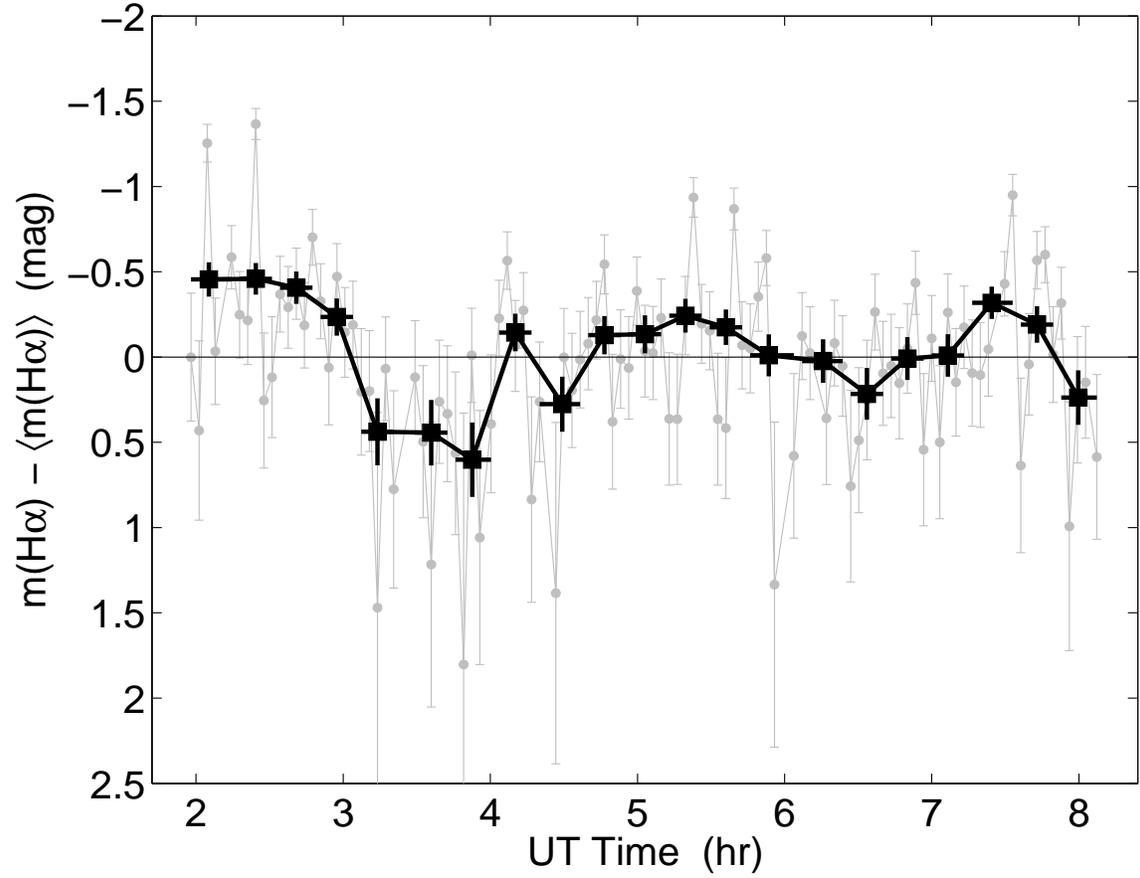,width=6.0in}}
\caption{H$\alpha$ light curve of BRI0021 from P60 narrow-band filter
observations.  The grey points are the raw light curve from the 180 s
exposures, while the black squares show the light curve binned to 900
s resolution.  The light curve is shown as magnitude fluctuations
relative to the mean brightness of the source, and thus provides an
indication of the H$\alpha$ variability timescale and amplitudes.  We
find variations of about a factor of 3 on $\sim 1$ hr timescales.
\label{fig:briha}}
\end{figure}

Finally, we observed BRI0021 with the robotic 60-inch telescope at
Palomar Observatory using an H$\alpha$ imaging filter with a width of
20 \AA.  We obtained a total of 106 images with an exposure time of
180 s each, and an on-source efficiency of $90\%$.  Since the purpose
of these observations was to search for H$\alpha$ variability we did
not obtain off-band images for absolute calibration.  Instead, we
constructed an H$\alpha$ light curve relative to five nearby stars.
We find that for the comparison stars the root-mean-square
fluctuations relative to their mean brightness is about $3\%$.  The
light curve of BRI0021, shown in Figure~\ref{fig:briha}, exhibits
clear variability on a $\sim 0.5-2$ hr timescale, with peak to trough
variations of about a factor of 3.

\section{Magnetic Activity Trends in Ultracool Dwarfs} 
\label{sec:prop}

\begin{figure}
\centerline{\psfig{file=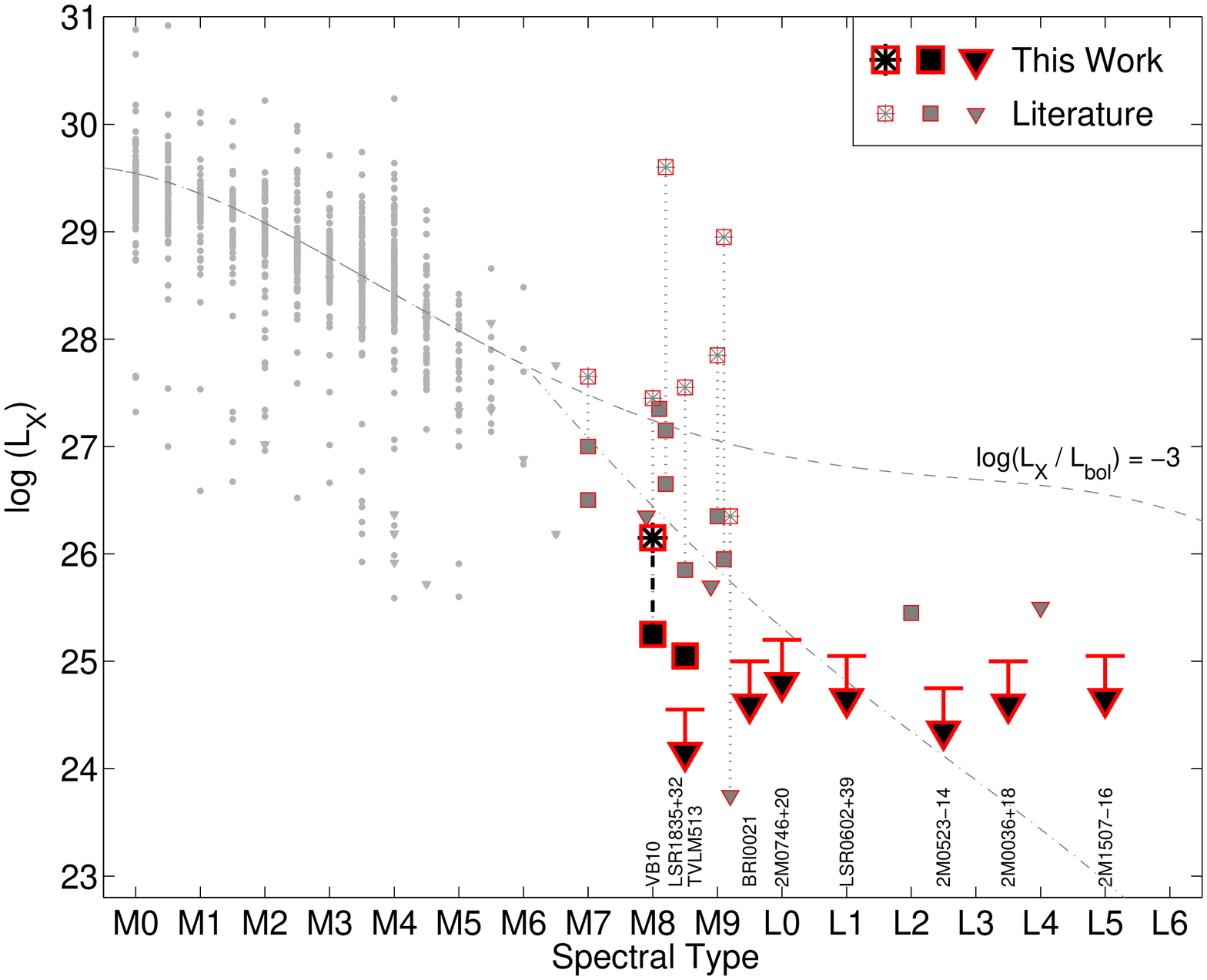,width=6.0in,angle=90}}
\caption{X-ray luminosity of M and L dwarfs plotted as a function of
spectral type.  For the ultracool dwarfs ($>{\rm M7}$), quiescent
emission is marked as squares, upper limits are marked as inverted
triangles, and flares are marked as asterisks.  Objects for which both
flares and quiescent emission have been observed are connected with
dotted lines.  The objects observed as part of our survey are noted at
the bottom of the figure.  The dashed line marks a constant ratio of
${\rm log}\,(L_X/L_{\rm bol})=-3$, the level of saturated emission in
early- and mid-M dwarfs, while the dot-dashed line represents the drop
in activity characterized by Equation~\ref{eqn:lx}.
\label{fig:lx}}
\end{figure}

\begin{figure}
\centerline{\psfig{file=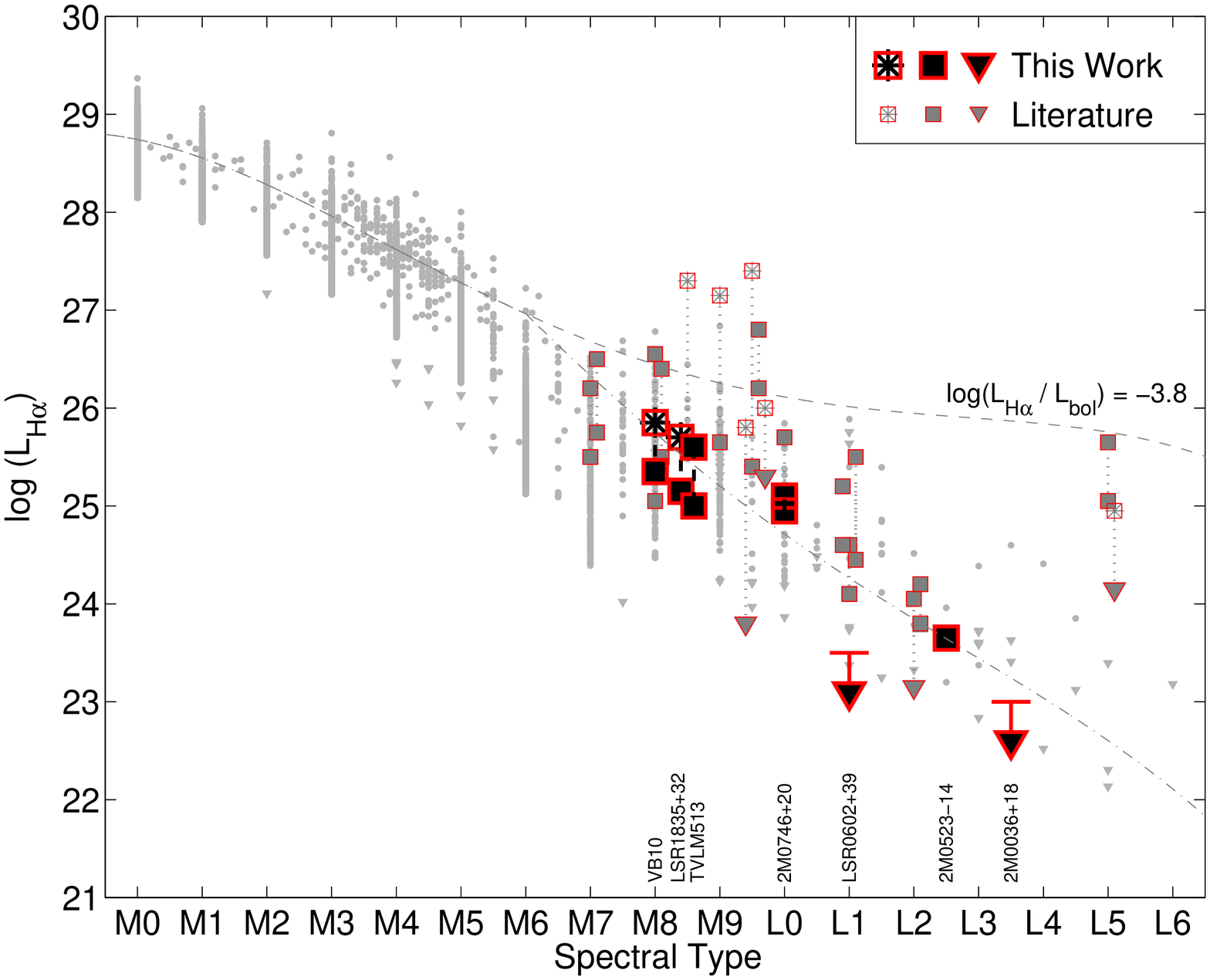,width=6.0in,angle=90}}
\caption{H$\alpha$ luminosity of M and L dwarfs plotted as a function
of spectral type.  For the ultracool dwarfs ($>{\rm M7}$), quiescent
emission is marked as squares, upper limits are marked as inverted
triangles, and flares are marked as asterisks.  Objects for which both
flares and quiescent emission have been observed are connected with
dotted lines.  The objects observed as part of our survey are noted at
the bottom of the figure.  The dashed line marks a constant ratio of
${\rm log}\,(L_{\rm rad}/L_{\rm bol})=-3.8$, the level of saturated
emission in early- and mid-M dwarfs, while the dot-dashed line
represents the drop in activity characterized by
Equation~\ref{eqn:lha}.
\label{fig:lha}}
\end{figure}

\begin{figure}
\centerline{\psfig{file=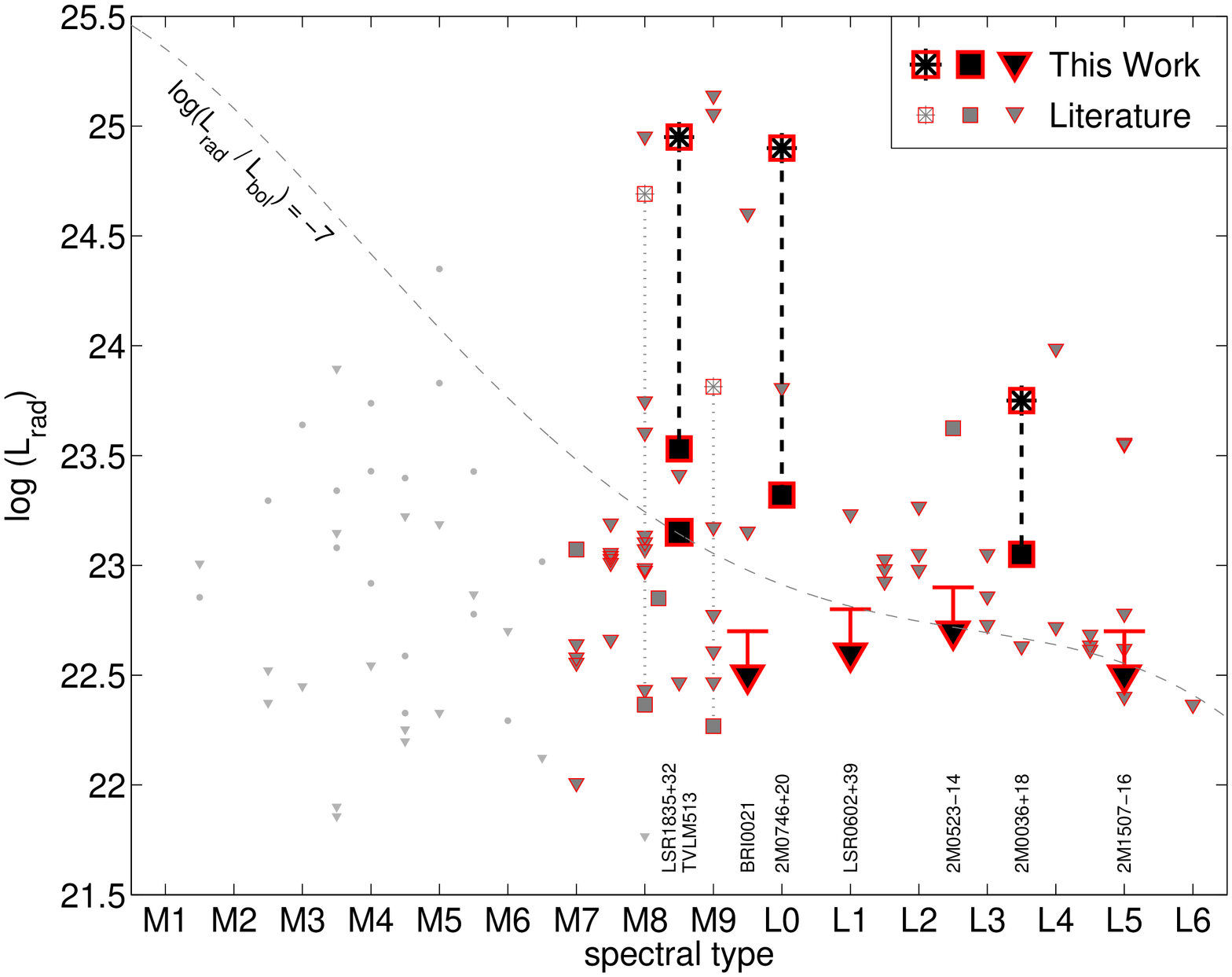,width=6.0in,angle=90}}
\caption{Radio luminosity of M and L dwarfs plotted as a function of
spectral type.  For the ultracool dwarfs ($>{\rm M7}$), quiescent
emission is marked as squares, upper limits are marked as inverted
triangles, and flares are marked as asterisks.  Objects for which both
flares and quiescent emission have been observed are connected with
dotted lines.  The objects observed as part of our survey are noted at
the bottom of the figure.  The dashed line marks a constant ratio of
${\rm log}\,(L_{\rm rad}/L_{\rm bol})=-7$, the approximate level of
radio activity in ultracool dwarfs.
\label{fig:lr}}
\end{figure}

\begin{figure}
\centerline{\psfig{file=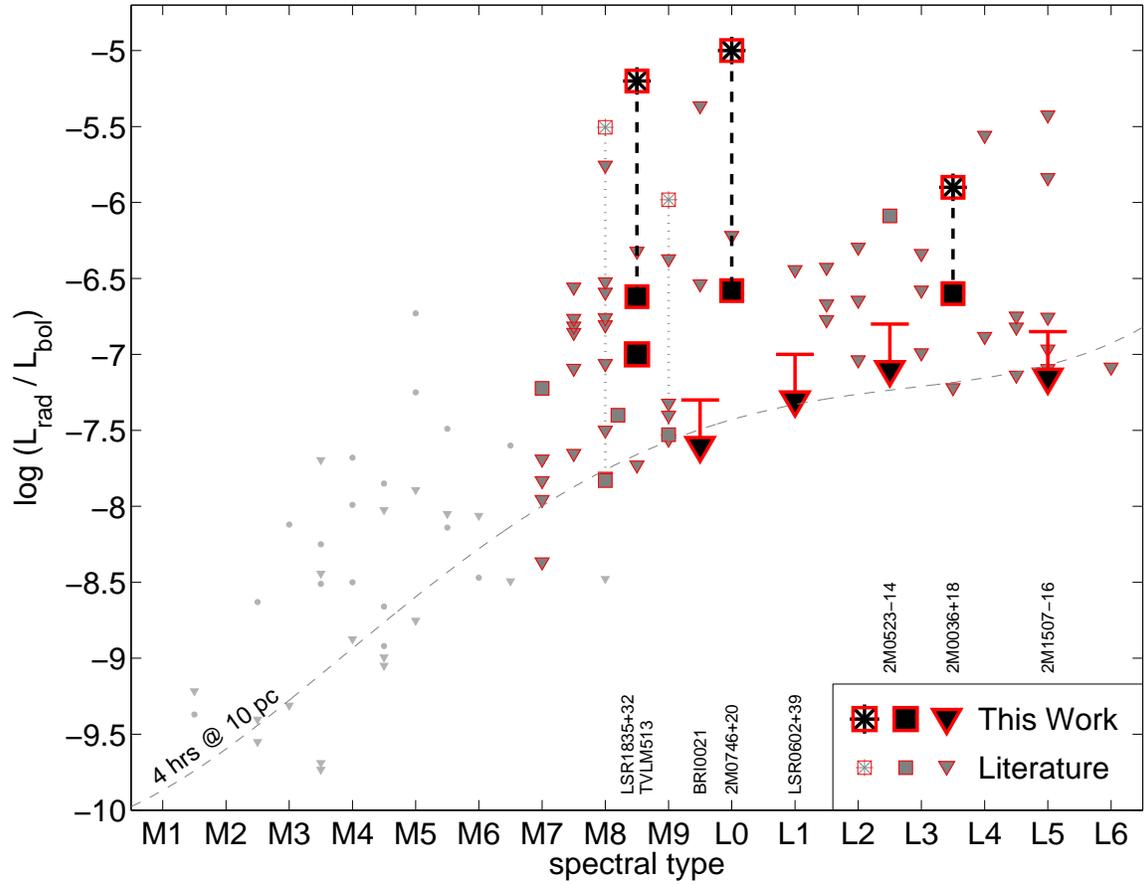,width=6.0in,angle=90}}
\caption{Ratio of radio luminosity to bolometric luminosity for M and
L dwarfs plotted as a function of spectral type.  Symbols are as in
Figure~\ref{fig:lr}.  The increase in $L_{\rm rad}/L_{\rm bol}$ as a
function of spectral type is clearly seen.  The dashed line indicates
the $3\sigma$ detection limit in a typical VLA observation for an
object located at a distance of 10 pc.
\label{fig:lrlbol}} 
\end{figure}

The results of our program, including the observations presented here,
in Papers I and II, in \citet{brr+05}, and in \citet{brp+09}, are
summarized in Figures~\ref{fig:lx}--\ref{fig:lrlbol}.  To supplement
these observations, we analyze unpublished archival {\it Chandra} and
VLA observations of the M8 dwarf LP\,412-31 in
Appendix~\ref{sec:app1}, and we collected all of the available
observations of late-M and L dwarfs from the literature.  These data
are summarized in Table~\ref{tab:all}.  To determine the activity
trends we also include comparison samples of early- and mid-M dwarfs.
Below we discuss the inferred activity trends for each of the main
indicators.

\subsection{X-rays}

Our survey provides a substantial increase in the number of ultracool
dwarfs observed in X-rays.  The X-ray luminosities of our sources, and
objects available in the literature, are plotted in
Figure~\ref{fig:lx}.  The typical quiescent X-ray luminosities range
from about $10^{29.5}$ erg s$^{-1}$ for M0 objects to about $10^{28}$
erg s$^{-1}$ for mid-M, and $10^{26.5}$ erg s$^{-1}$ for late-M
dwarfs.  The L dwarfs have typical limits of $\lesssim 10^{25}$ erg
s$^{-1}$.  The overall decline in X-ray luminosity is roughly
proportional to the decline in bolometric luminosity for spectral
types M0-M6, at a level of $L_X/L_{\rm bol} \approx 10^{-3}$.  This is
the well-known saturated X-ray emission in early- and mid-M dwarfs
\citep{pmm+03}.  The subsequent decline in X-ray luminosity, however,
is more rapid than expected from the same relation.

\begin{figure}
\centerline{\psfig{file=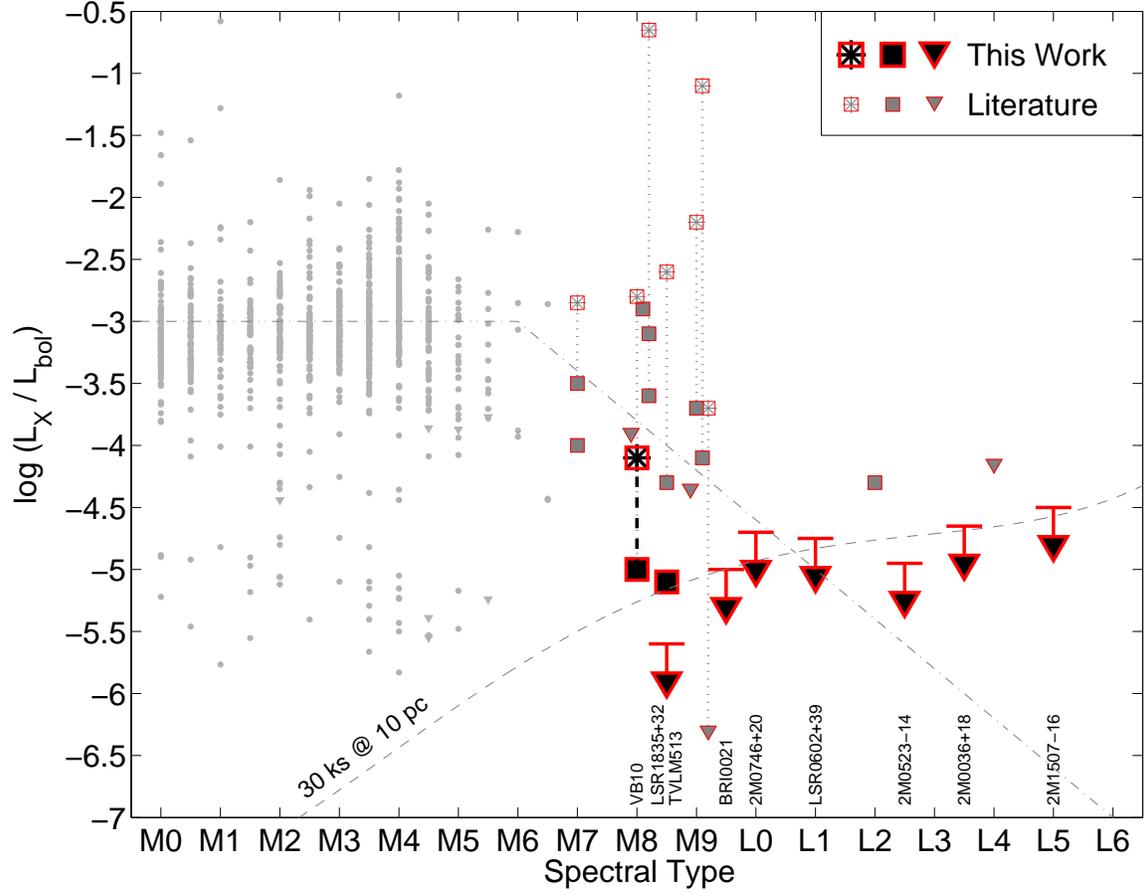,width=6.0in,angle=90}}
\caption{Ratio of X-ray to bolometric luminosity for M and L dwarfs
plotted as a function of spectral type.  Symbols are as in
Figure~\ref{fig:lx}.  The decline in X-ray activity beyond spectral
type of about M6 is clearly seen.  The dot-dashed line represents the
drop in activity characterized by Equation~\ref{eqn:lx}, while the
dashed line indicates the $3\sigma$ detection limit in a 30 ks {\it
Chandra} observation for an object located at a distance of 10 pc.
\label{fig:lxlbol}}
\end{figure}

To address this decline, and to assess the X-ray {\it activity} level
we plot the ratio of quiescent X-ray to bolometric luminosity for
spectral types M0-L5 in Figure~\ref{fig:lxlbol}.  This ratio exhibits
a clear trend of saturated emission ($L_X/L_{\rm bol}\approx 10^{-3}$)
down to a spectral type of about M6, followed by a significant decline
to $L_X/L_{\rm bol}\sim 10^{-4}$ for M7-M9, and $L_X/L_{\rm bol}
\lesssim 10^{-5}$ beyond M9.  With the exception of the very weak and
marginal detection of Kelu-1 (4 photons in 24 ks; \citealt{aob+07}),
no other L dwarf has been detected in the X-rays to date.  Our program
significantly extends the number of observed L dwarfs from 2 to 7, and
also provides the deepest available limits.  To quantify the decline
in X-ray activity we use a broken power law model:
\begin{equation}
{\rm log}\,\left(\frac{L_X}{L_{\rm bol}}\right)\approx
\begin{cases}
-3                        & {\rm for\,\,M<M6},\\
-3-0.4\times({\rm SP-M6}) & {\rm for\,\,M\ge M6},
\end{cases}
\label{eqn:lx}
\end{equation}
where SP is the spectral type.  This model accounts for the general
trend of declining quiescent X-ray emission (Figure~\ref{fig:lxlbol}).

Although the dynamic range for detectable quiescent emission in terms
of $L_X/L_{\rm bol}$ is more limited for ultracool dwarfs compared to
the early- and mid-M dwarfs, we find a dispersion of about two orders
of magnitude in the level of activity for spectral types M7-M9,
$L_X/L_{\rm bol}\approx -5$ to $-3$.  If we include non-detections,
the spread is at least three orders of magnitude.  For comparison,
early- and mid-M dwarfs exhibit an overall spread of nearly four
orders of magnitude.

Taking into account the detection of X-ray flares, we find that
objects with spectral types M7-M9 are capable of reaching values of
$L_X/L_{\rm bol}$ similar to, or even in excess of, early M dwarfs.
Moreover, we find that 7 of the 11 detected sources beyond spectral
type M7 produce flaring emission (or 7 out of the 21 observed
sources).  This suggests that a substantial fraction of the active
sources produce flares in addition to quiescent emission.  The typical
duration of the detected flares is $\sim 1$ hr, with a range of about
$10^2-10^4$ s.  Summing the total time which the M7-L5 sources spend
in a flaring state, we find that it represents about $6\%$ of the
total on-source exposure time.  Separating the sample into M7-M9 and L
dwarfs, we find that the former spend about $9\%$ of the time in
flares, while the latter spend $\lesssim 4\%$ of the time in a flare
state ($90\%$ confidence limit assuming a typical flare duration of 1
hr).  For our own sample of nine M8-L5 objects, which was observed in
a uniform manner with {\it Chandra}, we find a fraction of time in a
flaring state of only $3\%$.

\subsection{H$\alpha$}

The H$\alpha$ luminosities of our targets, and M and L dwarfs from the
literature, are shown in Figure~\ref{fig:lha}.  We find a continuous
decline from a luminosity of about $10^{28.5}$ erg s$^{-1}$ in early-M
dwarfs to about $10^{27}$ erg s$^{-1}$ in mid-M dwarfs, and about
$10^{25.5}$ erg s$^{-1}$ for late-M dwarfs.  Early-L dwarfs have a
typical H$\alpha$ luminosities of only $\sim 10^{24}$ erg s$^{-1}$.
As in the case of the X-ray emission, the overall decline in H$\alpha$
luminosity is roughly proportional to the decline in bolometric
luminosity for spectral types M0-M6, at a saturated level of $L_{\rm
H\alpha}/L_{\rm bol} \approx 10^{-3.8}$.  The subsequent decline,
however, is more rapid than expected from the same relation.  The
decline relative to the saturated level occurs at about the same
spectral type in both H$\alpha$ and X-rays.

\begin{figure}
\centerline{\psfig{file=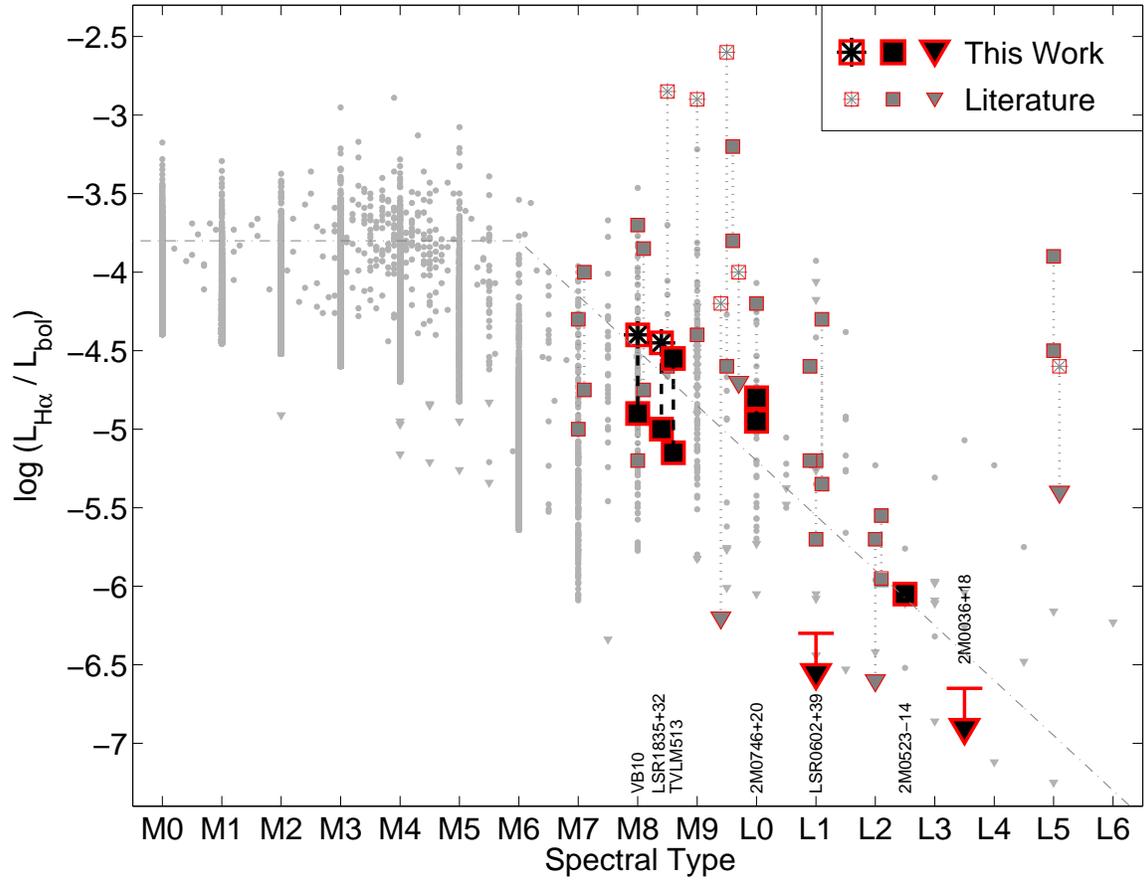,width=6.0in,angle=90}}
\caption{Ratio of H$\alpha$ to bolometric luminosity for M and L
dwarfs plotted as a function of spectral type.  Symbols are as in
Figure~\ref{fig:lha}.  The decline in H$\alpha$ activity beyond
spectral type of about M6 is clearly seen.  The dot-dashed line
represents the drop in activity characterized by
Equation~\ref{eqn:lha}.
\label{fig:lhalbol}}
\end{figure}

The ratio of H$\alpha$ to bolometric luminosity (i.e., H$\alpha$
activity) is shown in Figure~\ref{fig:lhalbol}.  The ratio is
relatively constant at a level of $L_{\rm H\alpha}/L_{\rm bol}\approx
10^{-3.8}$ down to about spectral type M6, and rapidly declines by at
least two orders of magnitude between M6 and L3.  The objects we
observed in this project exhibit the same trend, although some of our
limits are deeper than in past observations due to the long duration
of our observations.  As in the case of the X-ray emission, we
quantify the drop in H$\alpha$ activity using a broken power law:
\begin{equation}
{\rm log}\,\left(\frac{L_{\rm H\alpha}}{L_{\rm bol}}\right)\approx
\begin{cases}
-3.8                         & {\rm for\,\,M<M6},\\
-3.8-0.35\times({\rm SP-M6}) & {\rm for\,\,M\ge M6}.
\end{cases}
\label{eqn:lha}
\end{equation}
This model accounts for the general behavior of the quiescent
H$\alpha$ emission, and is in good agreement with the decline in X-ray
activity.  As in the case of the X-ray emission, the few detected
H$\alpha$ flares (dominated by objects with spectral types M7-M9) have
value of $L_{\rm H\alpha}/L_{\rm bol}$ similar to, or in excess of the
quiescent emission from early M dwarfs.  For M0-L5 objects, the
overall spread in $L_{\rm H\alpha}/L_{\rm bol}$ within each spectral
type bin is about two orders of magnitudes.

The fraction of M7-L5 objects which produce H$\alpha$ flares is
difficult to assess given the typical short duration of most
spectroscopic observations.  However, of the $\sim 660$ such objects
that we found in the literature (Figure~\ref{fig:lhalbol}), about 20
exhibit flares or significant variability, leading to a fraction of
about $3\%$.  This is a significantly smaller fraction than in the
X-rays ($\sim 30\%$).  The most likely explanation for this order of
magnitude discrepancy is a sample selection bias, namely X-ray
observations are expensive and have thus been focused on the nearest
and most likely to be active sources.  Conversely, H$\alpha$ flares
are likely to be under-reported, such that only observations of the
most extreme flares tend to be published.  If we restrict the analysis
of flare rates to our uniform sample of ultracool dwarfs, we find that
only 1/9 exhibit variable X-ray emission (1/2 of the detected
sources), while 5/8 exhibit variable H$\alpha$ emission (5/6 of the
detected sources).  Thus, it appears that in our sample, H$\alpha$
variability is more common than X-ray variability.

We note, however, that the typical flares/variability detected in our
observations are only a few times brighter than the quiescent
emission.  On the other hand, some of the flares reported in the
literature exceed the quiescent emission by up to two orders of
magnitude.  The lack of such flares in our observations places a limit
of $\lesssim 0.04$ per hour on their rate in the spectral type range
M8-L3 ($90\%$ confidence level).  Assuming a typical flare duration of
about 0.5 hr, the corresponding limit on the duty cycle is $\lesssim
2\%$.  This is comparable to recent estimates of flare duty cycles of
$\sim 5\%$ and $\sim 2\%$ for late-M and L dwarfs, respectively
\citep{scb+07}.

\subsection{Radio}

In the radio band, the observed trends in luminosity and $L_{\rm
rad}/L_{\rm bol}$ exhibit a remarkable contrast to the X-ray and
H$\alpha$ trends (see also \citealt{ber06}, Papers I and II, and
\citealt{aob+07}).  In particular, the detected sources with spectral
types M0-L5 exhibit nearly constant quiescent radio luminosity,
$L_{\rm rad}\approx 10^{23\pm 0.5}$ erg s$^{-1}$
(Figure~\ref{fig:lr}).  Radio flares (and pulsing emission;
\citealt{hbl+07,brp+09}) reach higher luminosities of up to $10^{25}$
erg s$^{-1}$.  At the present, no object beyond spectral type L3.5 has
been detected in the radio \citep{brr+05}.

The trend for $L_{\rm rad}/L_{\rm bol}$ naturally exhibits an
increase, from about $10^{-9}$ for early-M dwarfs, to $10^{-8}$ for
mid-M dwarfs, and about $10^{-6.5}$ for ultracool dwarfs.  The flaring
and pulsing emission reach $L_{\rm rad}/L_{\rm bol}\approx 10^{-5}$.
As in the case of the X-ray emission, and unlike in H$\alpha$, the
fraction of sources with detectable quiescent emission that also
produce flares or pulses is high, $\sim 50\%$.

The bulk of the non-detected ultracool dwarfs have limits that are
comparable in luminosity to the quiescent emission from the detected
sources.  This is simply a reflection of the limited dynamic range of
the existing radio observations, which generally consist of $\sim
1-10$ hr with the VLA.  The fractional detection rate may also be
influenced by source variability on long timescales
\citep{ber06,adh+07}.

\subsection{Breakdown of the Radio/X-ray Correlation}

Coronally active stars ranging from spectral type F to mid-M exhibit a
tight correlation between their radio and X-ray emission, with $L_{\rm
\nu,rad}/L_X\approx 10^{-15.5}$ Hz$^{-1}$ for \citep{gb93,bg94}.  The
first detection of radio emission from an ultracool dwarf, with
$L_{\rm \nu,rad}/ L_X\gtrsim 10^{-11.5}$ Hz$^{-1}$ \citep{bbb+01},
indicated that this correlation may be severely violated in these
objects.  Subsequent radio detections of ultracool dwarfs yielded
similar results \citep{ber02}.  Indeed, one of the main motivations
for our simultaneous radio/X-ray observations is to investigate the
nature of this breakdown.  Prior to our survey, most of the available
radio and X-ray observations of ultracool dwarfs were obtained years
apart, and could have been influenced by long-term variability.

\begin{figure}
\centerline{\psfig{file=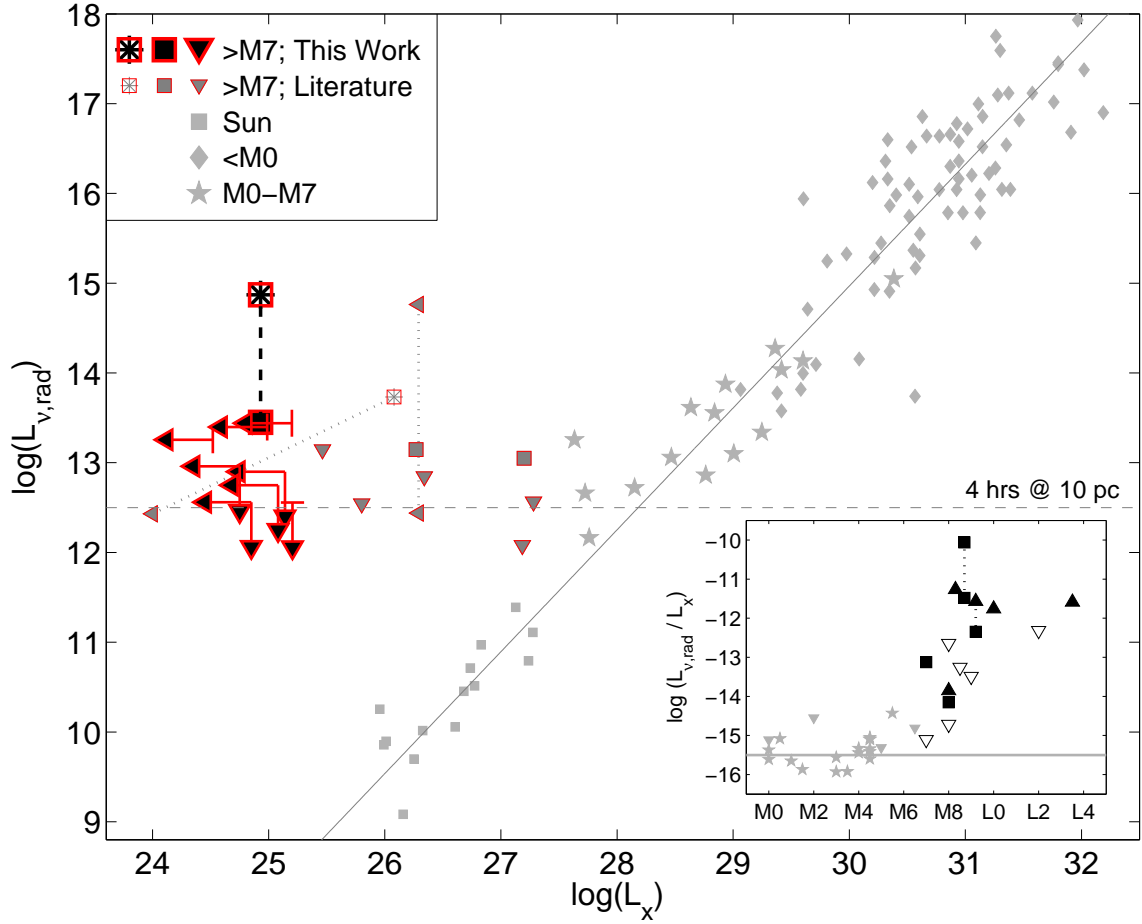,width=6.0in,angle=90}}
\caption{Radio luminosity plotted as function of X-ray luminosity for
a wide range of coronally active stars.  The radio/X-ray correlation
breaks down sharply, but smoothly, at about spectral type M7-M9.
\label{fig:gb}} 
\end{figure}

The results for all objects in our survey, supplemented by data from
the literature, are shown in Figure~\ref{fig:gb}.  The tight
correlation for spectral types earlier than M7 is evident, as is the
decline in both radio and X-ray luminosity as a function of spectral
type.  The best-fit linear trend is:
\begin{equation}
{\rm log}\,(L_{\rm \nu,rad})=1.36[{\rm log}\,(L_X)-18.97],
\label{eqn:lxlr}
\end{equation}
which in the range of X-ray luminosities for M0-M6 dwarfs, $L_X\sim
10^{28}-10^{29.5}$, corresponds to $L_{\rm \nu,rad}/L_X\approx
10^{-15.5}$ Hz$^{-1}$.  Beyond spectral type M6, however, we find a
clear trend of increasing ratio of $L_{\rm \nu,rad}/L_X$, from about
$10^{-14}$ for M7-M8 to $\gtrsim 10^{-12}$ beyond M9.  We note that
not all objects necessarily violate the correlation, since some are
detected in the X-rays with no corresponding radio emission.  However,
with the exception of the marginal detection of Kelu-1, all of these
objects have spectral types of M7-M9.

A comparison of the trends in Figures~\ref{fig:lx} and \ref{fig:lr}
suggests that the breakdown in the radio/X-ray correlation beyond
$\sim {\rm M7}$ is largely due to the rapid decline in quiescent X-ray
luminosity.  This is a surprising result since the strong correlation
for objects earlier than M6 suggests that heating (X-ray emission) and
particle acceleration (radio emission) are either directly correlated,
or share a common origin, presumably through dissipation of the
magnetic fields (e.g., \citealt{gb93}).  Since the radio luminosities
remain relatively unchanged from early-M to mid-L dwarfs, it appears
that the fraction of magnetic energy that goes into accelerating
electrons is roughly constant, and that the process of field
dissipation remains uniformly efficient despite the increasing
neutrality of the stellar atmospheres.  Atmospheric neutrality has
been argued to cause the decline in {\it
chromospheric}\footnotemark\footnotetext{The investigation of this
effect by \citet{mbs+02} does not directly pertain to X-ray and radio
emission since only physical conditions applicable to stellar
chromospheres were investigated.}  heating (i.e., H$\alpha$ emission)
due to the resulting decrease in magnetic stresses and field
dissipation \citep{mbs+02}.

Since the field dissipation and electron acceleration efficiency
remain relatively unchanged, we are forced to conclude that the
breakdown of the radio/X-ray correlation is due to a decrease in the
plasma heating efficiency.  This may be due to enhanced trapping of
the radio-emitting electrons, if these electrons are directly
responsible for plasma heating in higher mass stars.  Another
possibility is a decline in the bulk coronal density leading to a
strong suppression in X-ray heating, but with a minor impact on the
radio emission, which requires a smaller population of relativistic
electrons.  Alternatively, the geometry of the radio-emitting regions
may evolve to smaller sizes such that their impact on large-scale
coronal plasma heating decreases.  This latter explanation appears
less likely since rotationally-stable quiescent radio emission and
periodic H$\alpha$ emission from several ultracool dwarfs in our
sample point to large magnetic field covering fractions (e.g.,
\citealt{bbg+08,brp+09}).

\begin{figure}
\centerline{\psfig{file=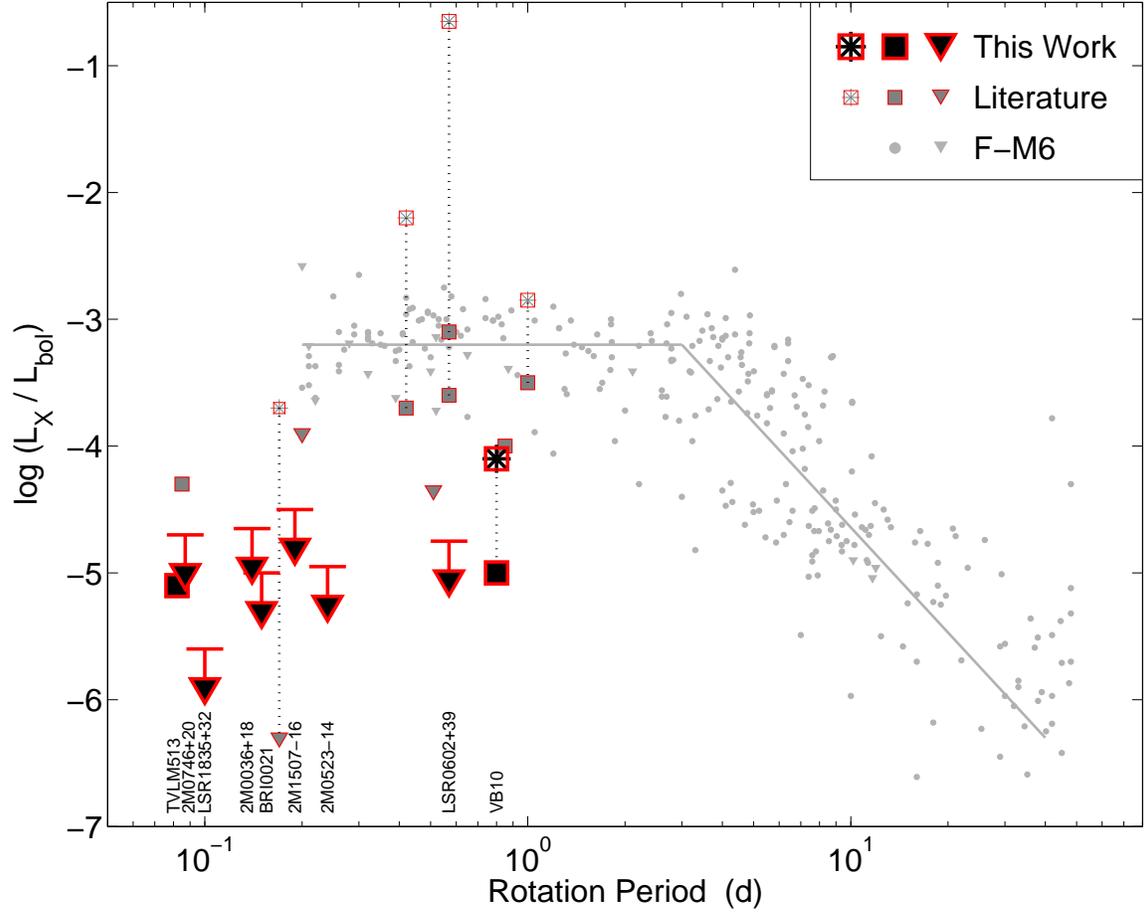,width=6.0in,angle=90}}
\caption{Ratio of X-ray to bolometric luminosity as a function of
rotation period for stars of spectral type F-L.  Symbols are as in
Figure~\ref{fig:lx}.  Data for the F-M6 dwarfs are from \citet{jjj+00}
and \citet{pmm+03}.  The median level of $L_X/L_{\rm bol}$ is about an
order of magnitude lower for $P<0.3$ d than for $P>0.3$ d, possibly as
a result of centrifugal stripping of the stellar corona.  The solid
line marks the trend of increasing in X-ray activity as a function of
rotation velocity, up to a saturation value of $L_X/L_{\rm bol}\approx
10^{-3}$.
\label{fig:lxlbol_rot}}
\end{figure}

Finally, the combination of rapid rotation and the shrinking
co-rotation radius of lower mass stars may lead to a decline in X-ray
emission through centrifugal stripping of the corona.  In
Figure~\ref{fig:lxlbol_rot} we plot the ratio $L_X/L_{\rm bol}$ as a
function of rotation period.  We find a decline in $L_X/L_{\rm bol}$
as a function of decreasing period for objects later than M7, with a
median value of $L_X/L_{\rm bol}\approx 10^{-4}$ for $P>0.3$ d, and
$L_X/L_{\rm bol}\approx 10^{-5}$ for $P<0.3$ d (for typical
parameters, $P\approx 0.3$ d corresponds to $v\sim 15-20$ km
s$^{-1}$).  For this decline to be due to coronal stripping, the
magnetic field scale height should be a few stellar radii.

\section{Summary and Future Prospects}
\label{sec:conc}

We presented simultaneous X-ray, radio, and H$\alpha$ observations of
three late-M and L dwarfs, which along with our previous published
results double the number ultracool dwarfs and triple the number of L
dwarfs observed in X-rays.  The overall X-ray detection fraction in
our survey, to a typical limit of $L_X/L_{\rm bol}\approx 10^{-5}$, is
about $20\%$.  Combining our sources with all objects later than M7
from the literature leads to a detection fraction of about $50\%$.
This fraction is dominated by late-M dwarfs, with at most $\sim 15\%$
of L dwarfs detected to date (this is based on the marginal detection
of Kelu-1).  We further find a significant drop in X-ray activity
beyond spectral type $\sim {\rm M7}$, to $L_X/L_{\rm bol} \approx
10^{-4}$ for M7-M9 and $L_X/L_{\rm bol}\lesssim 10^{-5}$ for L dwarfs.
The decline in X-ray activity, and the observed level of emission, are
similar to those measured for low mass stars and brown dwarfs in young
star forming regions \citep{fbg+02,ms02,pz02,gbg+07}.

A similar decline in activity is observed in H$\alpha$, although
$75\%$ of our targets are detected.  In the radio band, however, we
find that $L_{\rm rad}$ remains relatively unchanged in the range
M0-L4, and hence $L_{\rm rad}/L_{\rm bol}$ increases by about two
orders of magnitudes over the same spectral type range.  The rapid
decline in X-ray emission and the uniform level of radio emission lead
to a breakdown in the radio/X-ray correlation.  With our increased
sample of objects we find that the ratio transitions sharply, but
smoothly, from $L_{\rm \nu,rad}/ L_X\approx 10^{-15.5}$ to $\sim
10^{-11.5}$ Hz$^{-1}$ over the spectral type range M7-M9.  The
radio/X-ray correlation in objects earlier than M7 is likely due to a
correlated or common origin for particle acceleration and plasma
heating.  We conclude that its breakdown is due to a decline in the
plasma heating efficiency, as a result of efficient electron trapping,
increased neutrality, decrease in coronal density, a decline in the
size of the radio-emitting regions, or coronal stripping.

In light of our results, we suggest that continued observational
progress in the study of X-ray emission from ultracool dwarfs can be
achieved with two distinct approaches: (i) deeper observations of a
small number of nearby objects; and (ii) a shallower survey of a
significantly larger sample of objects than is currently available.
In the context of the former approach, our existing survey indicates
that most L dwarfs do not produce quiescent X-ray emission at a level
of $L_X/L_{\rm bol}\gtrsim 10^{-5}$ based on 30 ks observations.
Since our targets were located within about 15 pc, deeper limits can
only be achieved with $\sim 100$ ks observations of targets within
$\sim 5$ pc.  However, the only early-L dwarfs that come close to
matching this criterion (2M0036+18 and 2M1507$-$16) were already
observed as part of our survey.  Thus, it is unlikely that we can
observe any ultracool dwarfs to limits that are more than a factor of
$2-3$ times deeper than our current survey, and we therefore conclude
that {\it blind} long X-ray observations do not provide the best way
forward.  Instead, such observations should be targeted at objects
with strong radio and/or H$\alpha$ emission, for which the likelihood
of an X-ray detection may be higher.

In addition, a large shallow survey may be highly fruitful in
detecting X-ray emission from ultracool dwarfs.  The existing
observations indicate that late-M dwarfs are capable of producing
flares at a level similar to the saturated X-ray emission from early-M
dwarfs, $L_X/L_{\rm bol}\sim 10^{-3}$.  Such flares can be detected
with {\it Chandra} or {\it XMM-Newton} to a distance of $\sim 50$ pc
in about 10 ks.  Thus, a survey of about $50$ ultracool dwarfs within
this distance would require a similar amount of time to the existing
observations of ultracool dwarfs, but it will significantly increase
the chance of detecting rare flaring objects.  Coupled with
simultaneous radio observations, this survey will also allow us to
better assess the radio/X-ray correlation and its breakdown.

Beyond additional observations, progress in our understanding of the
decline in X-ray emission and the breakdown of the radio/X-ray
correlation requires detailed theoretical studies of coronal
conditions in ultracool dwarfs.  The existing work \citep{mbs+02}
investigated atmospheric conditions that are only applicable to the
stellar chromosphere, and led to the conclusion that dissipation of
the magnetic field is suppressed by the increasing neutrality of the
stellar atmospheres.  However, particle acceleration in the magnetic
fields appears to remain efficient at least to mid-L dwarfs as
evidenced by radio observations, suggesting that in the corona the
magnetic fields may couple effectively to the tenuous atmosphere.

\acknowledgements We thank the Chandra, Gemini, and VLA schedulers for
their invaluable help in coordinating these observations.  This work
has made use of the SIMBAD database, operated at CDS, Strasbourg,
France.  It is based in part on observations obtained at the Gemini
Observatory, which is operated by the Association of Universities for
Research in Astronomy, Inc., under a cooperative agreement with the
NSF on behalf of the Gemini partnership: the National Science
Foundation (United States), the Science and Technology Facilities
Council (United Kingdom), the National Research Council (Canada),
CONICYT (Chile), the Australian Research Council (Australia), CNPq
(Brazil) and CONICET (Argentina).  Data from the UVOT instrument on
Swift were used in this work.  Support for this work was provided by
the National Aeronautics and Space Administration through Chandra
Award Number G08-9013A issued by the Chandra X-ray Observatory Center,
which is operated by the Smithsonian Astrophysical Observatory for and
on behalf of the National Aeronautics Space Administration under
contract NAS8-03060.

\appendix  

\section{X-ray and Radio Observations of LP\,412-31}  
\label{sec:app1}

In addition to the observations presented in this paper, we retrieved
from the {\it Chandra} archive observations of the M8 dwarf LP\,412-31
that were obtained as part of program 09200198 (PI: Stelzer) on 2007
December 22.95 UT for a total of 40.6 ks.  This object is located at a
distance of $14.6\pm 0.1$ pc \citep{rc02}, and has a bolometric
luminosity of $L_{\rm bol}\approx 10^{-3.26}$ L$_\odot$.  It was
previously detected with {\it XMM-Newton} with a quiescent luminosity
of $L_X\approx 1.6\times 10^{27}$ erg s$^{-1}$, and a flare that
peaked at $L_X\approx 4.6\times 10^{29}$ erg s$^{-1}$ \citep{ssm+06}.

\begin{figure}
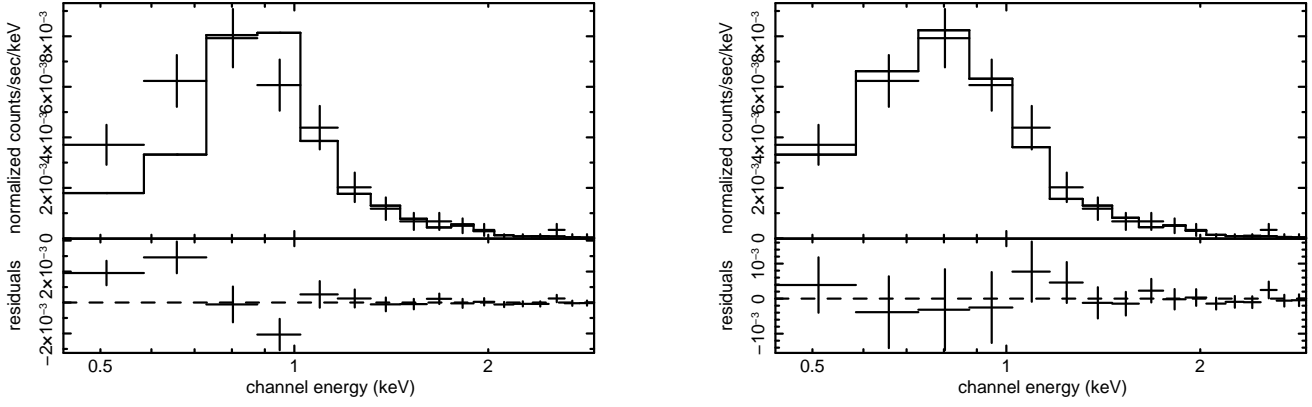

\centerline{\psfig{file=fig10a.ps,width=3.5in,angle=270}
\hspace{0.2in}\psfig{file=fig10b.ps,width=3.5in,angle=270}}
\caption{X-ray spectrum of LP\,412-31 from {\it Chandra} observations
(\S\ref{sec:app1}).  {\it Left:} A single-temperature Raymond-Smith
model provides a poor fit to the data.  {\it Right:} A two-temperature
Raymond-Smith model provides an excellent fit to the data with
resulting values of $kT\approx 0.3$ and 1 keV.
\label{fig:lp412-31_x}}
\end{figure}

We analyzed the data using CIAO version 3.4, and extracted counts in a
$3''$ radius circle centered on the position of LP\,412-31.  The
source is clearly detected, with a total of 239 counts in the $0.2-5$
keV range, compared to about 1 count expected from the background.  To
determine the X-ray flux we fit the spectrum using {\tt xspec} V11.
We restrict the fit to the energy range $0.3-3$ keV over which most of
the counts are detected.  Using a single-temperature Raymond-Smith
model with the abundance set to 0.3, we find a poor fit to the data
($\chi^2=22$ for 15 degrees of freedom) with a temperature of
$kT\approx 0.8$ keV (Figure~\ref{fig:lp412-31_x}a).  A significantly
improved fit ($\chi^2=5.5$ for 13 degrees of freedom) is obtained with
a two-temperature Raymond-Smith model, with temperatures of $kT\approx
0.3$ and $\approx 1$ keV (Figure~\ref{fig:lp412-31_x}b).  The
resulting flux of LP\,412-31 for this model is $2.0\times 10^{-14}$
erg cm$^{-2}$ s$^{-1}$ in the $0.3-3$ keV energy range (and about
$15\%$ lower for the $0.5-3$ keV energy range).  Thus, the X-ray
luminosity is $5.1\times 10^{26}$ erg s$^{-1}$, and the ratio relative
to the bolometric luminosity is $L_X/L_{\rm bol}\approx 10^{-3.6}$.
This value is about a factor of 3 times lower than the quiescent
luminosity measured with {\it XMM-Newton} \citep{ssm+06}.

LP\,412-31 was also observed with the VLA at 8.46 GHz as part of
program AS879 on four separate occasions (2006 October 31 and 2006
November 5, 7, and 9 for a total of about 13.4 hours), and as part of
program S90198 simultaneously with the {\it Chandra} observation.  We
obtained the public data from the archive and processed the
observations in the manner described in \S\ref{sec:rad}.  The combined
2006 observations reveal a possible weak source coincident with the
position of LP\,412-31, with a flux of $F_\nu=37\pm 11$ $\mu$Jy (using
only the two longest observations, which account for $75\%$ of the
data, we find $F_\nu=45\pm 12$ $\mu$Jy).  In the subsequent
simultaneous observation we do not detect any emission at the position
of LP\,412-31, with a $3\sigma$ upper limit of $F_\nu\lesssim 25$
$\mu$Jy.  This indicates that the source either fluctuated in
brightness by at least a factor of 2, or that the initial weak source
was spurious.  The resulting limit on the radio luminosity is $L_{\rm
\nu,rad}\lesssim 6.4\times 10^{12}$ erg s$^{-1}$ Hz$^{-1}$.

Combining the X-ray and radio results, we find a limit of $L_{\rm
\nu,rad}/L_X\lesssim 10^{-14.4}$ Hz$^{-1}$, about a factor of 40 times
higher than the expected value (Equation~\ref{eqn:lxlr}).


\clearpage
\begin{deluxetable}{lccccccl}
\tabletypesize{\scriptsize}
\tablecolumns{8}
\tabcolsep0.05in\footnotesize
\tablewidth{0pc}
\tablecaption{Target Properties
\label{tab:sources}}
\tablehead{
\colhead{Source}              &
\colhead{Sp.~Type}            &
\colhead{$d$}                 &
\colhead{${\rm log}(L_{\rm bol}/L_\odot$)} &
\colhead{$v{\rm sin}i$}       &
\colhead{${\rm log}(L_{\rm H\alpha}/L_{\rm bol})$}   &
\colhead{${\rm log}(\nu_{\rm rad}L_{\rm \nu,rad}/L_{\rm bol})$} &
\colhead{Ref.}               \\
\colhead{}                    &
\colhead{}                    &
\colhead{(pc)}                &
\colhead{}                    &
\colhead{(km s$^{-1}$)}       &
\colhead{}                    &
\colhead{}                    &
\colhead{}                 
}
\startdata
BRI0021     & M9.5 & 11.5 & $-3.44$ & 34 & $<-6.0$ \to\ $-4.2$  & $-7.08\pm 0.09$     & 1-4   \\
LSR0602+39  & L1   & 10.6 & $-3.73$ & 9  & $<-6.0$            & $<-7.32$            & 4-6   \\
2M0523$-$14 & L2.5 & 13.4 & $-3.78$ & 21 & $-6.5$             & $<-6.95$ \to\ $-6.18$ & 4,6-8 
\enddata
\tablecomments{$^a$ Limits are $3\sigma$.\\
References: [1] \citet{bm95}; [2] \citet{rkg+99}; [3] \citet{mb03}; 
[4] \citet{ber02}; [5] \citet{slr+03}; [6] \citet{rb08}; 
[7] \citet{scb+07}; [8] \citet{adh+07}.}
\end{deluxetable}

\begin{deluxetable}{lcccccclc}
\tabletypesize{\scriptsize}
\tablecolumns{9}
\tabcolsep0.05in\footnotesize
\tablewidth{0pc}
\tablecaption{Radio Observations and Results
\label{tab:radio}}
\tablehead{
\colhead{Source}              &
\colhead{Date}                &
\colhead{Time}                &
\colhead{Exposure}            &
\colhead{Phase Cal.}          &
\colhead{Cal.~Exposure}       &
\colhead{$\nu$}               &
\colhead{$F_\nu\,^a$}         &
\colhead{${\rm log}(\nu_{\rm rad}L_{\rm \nu,rad}/L_{\rm bol})\,^a$} \\
\colhead{}                    &
\colhead{(UT)}                &
\colhead{(UT)}                &
\colhead{(s)}                 &
\colhead{}                    &
\colhead{(s)}                 &
\colhead{(GHz)}               &
\colhead{($\mu$Jy)}           &
\colhead{}                    
}
\startdata
BRI0021     & 2008 Oct 29 & 00:41:03 -- 09:38:38 & 220 & J0016$-$002 & 40 & 4.86 & $<54$  & $<-7.52$ \\
            &             &                      &     &             &    & 8.46 & $<48$  & $<-7.33$ \\
LSR0602+39  & 2008 Jan 18 & 01:02:40 -- 09:22:30 & 270 & J0555+398   & 30 & 8.46 & $<42$  & $<-7.17$ \\
2M0523$-$14 & 2007 Dec 30 & 02:16:00 -- 10:07:05 & 280 & J0539$-$158 & 35 & 4.86 & $<105$ & $<-6.76$ \\
            &             &                      &     &             &    & 8.46 & $<42$  & $<-6.92$ 
\enddata
\tablecomments{$^a$ Limits are $3\sigma$.}
\end{deluxetable}

\begin{deluxetable}{lccccc}
\tabletypesize{\scriptsize}
\tablecolumns{6}
\tabcolsep0.05in\footnotesize
\tablewidth{0pc}
\tablecaption{X-ray Observations and Results
\label{tab:xray}}
\tablehead{
\colhead{Source}              &
\colhead{Date}                &
\colhead{Time}                &
\colhead{Exposure}            &
\colhead{$F_X$}               &
\colhead{${\rm log}(L_X/L_{\rm bol})$}   \\
\colhead{}                    &
\colhead{(UT)}                &
\colhead{(UT)}                &
\colhead{(s)}                 &
\colhead{(erg cm$^{-2}$ s$^{-1}$)} &
\colhead{}                 
}
\startdata
BRI0021     & 2008 Oct 29 & 02:44:02 -- 11:33:31 & 29797 & $<8.0\times 10^{-16}$ & $<-5.04$  \\
LSR0602+39  & 2008 Jan 18 & 00:27:20 -- 09:28:50 & 29664 & $<8.0\times 10^{-16}$ & $<-4.82$  \\
2M0523$-$14 & 2007 Dec 30 & 02:02:46 -- 11:01:21 & 29775 & $<8.0\times 10^{-16}$ & $<-4.57$  
\enddata
\tablecomments{$^a$ Limits are $95\%$ confidence level.}
\end{deluxetable}

\begin{deluxetable}{lcccccc}
\tabletypesize{\scriptsize}
\tablecolumns{7}
\tabcolsep0.05in\footnotesize
\tablewidth{0pc}
\tablecaption{H$\alpha$ Observations and Results
\label{tab:ha}}
\tablehead{
\colhead{Source}              &
\colhead{Date}                &
\colhead{Time}                &
\colhead{Telescope}           &
\colhead{Exposure}            &
\colhead{EW}                  &
\colhead{${\rm log}(L_{\rm H\alpha}/L_{\rm bol})$}   \\
\colhead{}                    &
\colhead{(UT)}                &
\colhead{(UT)}                &
\colhead{}                    &
\colhead{(s)}                 &
\colhead{(\AA)}               &
\colhead{}                 
}
\startdata
BRI0021     & 2008 Oct 29 & 01:57:58 -- 08:10:19 & Gemini-South    & $73\times 300$  & $1.00\pm 0.05$ & $-6.05$ \\
LSR0602+39  & 2008 Jan 18 & 01:13:28 -- 09:49:24 & ARC 3.5-m       & $38\times 600$  & $<0.25$        & $<-6.3$ \\
2M0523$-$14 & 2007 Dec 30 & 02:02:46 -- 11:01:21 & Palomar 60-inch & $106\times 180$ & \nod$\,^a$     & \nod  
\enddata
\tablecomments{$^a$ We did not calibrate the H$\alpha$ flux; the
purpose of the observations was to search for variability.}
\end{deluxetable}

\clearpage
\begin{deluxetable}{llllllll}
\tabletypesize{\scriptsize}
\tablecolumns{8}
\tabcolsep0.03in\footnotesize
\tablewidth{0pc}
\tablecaption{Magnetic Activity Results from our Full Sample of 
Simultaneous Observations
\label{tab:all}}
\tablehead{
\colhead{Source}              &
\colhead{Sp.~Type}            &
\colhead{${\rm log}(L_X/L_{\rm bol})$} &
\colhead{${\rm log}(L_{\rm H\alpha}/L_{\rm bol})$} &
\colhead{${\rm log}(\nu_{\rm rad}L_{\rm \nu,rad}/L_{\rm bol})$} &
\colhead{${\rm log}(L_{\rm H\alpha}/L_X)\,^a$} &
\colhead{${\rm log}(\nu_{\rm rad}L_{\rm \nu,rad}/L_X)\,^b$} &
\colhead{Ref.}               \\
\colhead{}                    &
\colhead{}                    &
\colhead{}                    &
\colhead{}                    &
\colhead{}                    &
\colhead{}                    &
\colhead{}                    &
\colhead{}                 
}
\startdata
VB10        & M8       & $-5.0$ / $-4.1$(f)$\,^c$ & $-4.9$ / $-4.4$(f)    & [$<-7.8$]$\,^d$    & $-0.3$ / $0.1$(f) & [$<-2.8$]       & Paper II \\
LSR1835	    & M8.5     & $<-5.6$     	          & $-5.0$ / $-4.45$(f)   & $-7.0$	       & $>0.6$            & $>-1.4$         & Paper II \\
TVLM513	    & M8.5     & $-5.1$                   & $-5.15$ / $-4.55$     & $-6.6$ / $-5.2$(f) & $-0.05$ / $0.55$  & $-1.4$ / $-0.1$ & Paper I \\
BRI0021	    & M9.5     & $<-5.0$	          & [$<-6.2$ / $-4.2$(f)] & $<-7.3$	       & [$>0.8$(f)]       & \nod            & This Paper \\
2M0746+20   & L0$\,^e$ & $<-4.7$	          & $-4.95$ / $-4.8$      & $-6.6$ / $-5.0$(f) & $>-0.25$          & $>-1.9$         & 1 \\
LSR0602+39  & L1       & $<-4.75$	          & $<-6.3$		  & $<-7.0$	       & \nod              & \nod            & This Paper \\
2M0523$-$14 & L2.5     & $<-4.95$	          & $-6.05$		  & $<-6.8$	       & $>-1.1$           & \nod            & This Paper \\
2M0036+18   & L3.5     & $<-4.65$	          & $<-6.65$	          & $-6.6$ / $-5.9$(f) & \nod              & $>-1.95$        & 2 \\
2M1507$-$16 & L5       & $<-4.5$                  & [$<-5.8$]             & [$<-6.85$]         & \nod              & \nod            & 2 \\\hline\\
\multicolumn{8}{c}{\normalsize Additional Observations from the Literature} \\\hline
VB8         & M7      & $-3.5$ / $-2.85$(f) & $-5.0$ / $-2.85$(f) & $<-8.4$            & $-1.5$             & $<-4.9$             & 3,4 \\
LHS3003     & M7      & $-4.0$              & $-4.75$ / $-3.85$   & $-7.5$             & $-0.75$ / $0.15$   & $-3.5$              & 3,5,6 \\
DENIS1048   & M8      & $<-3.9$             & \nod                & $-7.8$ / $-5.5$(f) & \nod               & $>-3.9$             & 6,7 \\ 
Gl234b      & M8      & $-2.9$              & \nod                & $<-7.9$            & \nod               & $<-5.0$             & 3,8 \\
LP412-31    & M8      & $-3.6$ / $-0.65$(f) & $-4.0$              & $<-7.6$            & $-0.4$             & $<-4.0$             & 9-11, This Paper \\
Gl569B      & M8.5    & $-4.3$ / $-2.6$     & \nod                & $<-7.7$            & \nod               & $<-3.4$             & 8,12 \\
LHS2924     & M9      & $<-4.35$            & $-4.7$              & \nod               & $>-0.35$           & \nod                & 3 \\
LHS2065     & M9      & $-3.7$ / $-2.5$(f)  & $-4.4$ / $-2.9$(f)  & $<-7.3$            & $-0.7$ / $-0.4$(f) & $<-3.6$             & 8,13 \\
1RXS1159    & M9      & $-4.1$ / $-1.1$(f)  & \nod                & \nod               & \nod               & \nod                & 14,15 \\
LP944-20    & M9      & $<-6.3$ / $-3.7$(f) & $-5.3$              & $-7.5$ / $-6.0$(f) & $>1.0$             & $-2.3$(f) / $>-1.2$ & 16-18 \\
Kelu-1      & L2      & $-4.3$              & $-5.2$              & $<-7.0$            & $-0.9$             & $<-2.7$             & 19 \\
HD130948BC$\,^f$ & L4 & $<-4.15$            & \nod                & \nod               & \nod               & \nod                & 20               

\enddata \tablecomments{$^a$ The typical value for early and mid M
dwarfs is ${\rm log}(L_{\rm H\alpha}/L_X)\approx -0.5$
\citep{hj03}. \\ $^b$ The expected value at 8.46 GHz based on the
radio/X-ray correlation is ${\rm log}(\nu_{\rm rad}L_{\rm
\nu,rad}/L_X)\approx -5.6$. \\ $^c$ ``(f)'' designates flaring
emission. \\ $^d$ Values in brackets are from non-simultaneous
observations.\\ $^e$ 2M0746+20 is a binary system, and we assume that
the observed emission is due to the primary. \\ References: [1]
\citet{brp+09}; [2] \citet{brr+05}; [3] \citet{sfg95}; [4]
\citet{kll99}; [5] \citet{ma01}; [6] \citet{bp05}; [7] \citet{sl04};
[8] \citet{ber06}; [9] \citet{ssm+06}; [10] \citet{mb03}; [11]
\citet{scb+07}; [12] \citet{ste04}; [13] \citet{rs08}; [14]
\citet{hss+04}; [15] \citet{rs09}; [16] \citet{rbm+00}; [17]
\citet{mb02}; [18] \citet{bbb+01}; [19] \citet{aob+07}; [20]
\citet{smf+06}.}
\end{deluxetable}

\clearpage

\clearpage

\clearpage

\clearpage

\clearpage

\clearpage

\clearpage

\clearpage

\clearpage

\clearpage

\end{document}